\documentstyle[12pt,world_sci,cite,epsf,epsfig]{article}
\newskip\humongous \humongous=0pt plus 1000pt minus 1000pt

\newif\ifdtup

\def\Im{\mathop{\rm Im}}

\def\beq{\begin{equation}}
\def\eeq{\end{equation}}

\def\beqn{\begin{eqnarray}}
\def\eeqn{\end{eqnarray}}
\relax

\def\dotx{\dotx{\dot\overline{x}}}

\relax

\newcommand{\kpinn}{$K^{+}\rightarrow\pi^{+}\nu\overline{\nu}\;$}

\newcommand{\kpitwo}{$K^{+}\rightarrow\pi^{+}\pi^{0}\;$}

\begin{document}
~~~~~~~~~~~~~~~~~~~~~~~~~~~~~~~~~~~~~~~~~~~~~~~~~~~~~~~~~~~~~~~~~~~~~~~~~~~~~~~~~~~~{\bf BNL-65188}

\title{BNL Future Plans\footnotemark}
\author{ Laurence Littenberg \\
\it Physics Department, Brookhaven National Laboratory \\
 Upton, New York 11973 USA \\}
\maketitle
\section*{Abstract}

I discuss the prospects for a fixed target physics program at the
AGS in the RHIC era.  
\footnotetext{To be published in the proceedings of the KEK {\it Workshop on
Kaon, Muon, Neutrino Physics and Future} \\ 31 Oct - 1 Nov 1997.}
\thispagestyle{empty}
\section{Introduction}
\label{s: intro}

	In 1999, after almost 40 years of independent existence, the
Brookhaven Alternating Gradient Synchrotron (AGS) is scheduled to be
pressed into service as an injector to the Relativistic Heavy Ion
Collider (RHIC).  Although at first sight this seems like the end of
an era, in actuality, it represents a very attractive new opportunity.
For the AGS is actually needed by RHIC for only a few hours per day.
The balance of the time it is available for extracted proton beam work at
a very small incremental cost.  This represents the reverse of the
current situation in which the nuclear physics program gets access to
the AGS (for fixed target heavy ion experiments) at incremental cost,
while the base cost of maintaining the accelerator is borne by the
high energy physics program.

	Retaining the AGS for particle physics work would broaden the
US HEP program considerably, allowing continued exploitation of the
world's most intense source of medium energy protons.  As will be
discussed below, there are some very compelling experiments that can
best be done at the AGS.  These include a determination of the
Cabibbo-Kobayashi-Maskawa (CKM) matrix element $V_{td}$, probes of
Standard Model (SM) and non-SM CP violation, and low energy
manifestations of supersymmetry (SUSY).

\section{AGS Upgrades}
\label{s: agsu}

\begin{table}[htb]
\centering
\caption{Recent performance of the AGS}
\begin{tabular}{lllll} \hline
 &  1994 &  1995 & 1996 & 1997 \\ \hline
{\bf Proton Beams:} & & \cr
Beam Energy & 24 GeV & 24 GeV & 24 GeV & 24 GeV \\
Peak Beam Intensity (ppp) & $40 \times 10^{12}$ & $63 \times 10^{12}$ & $62 \times 10^{12}$ & $62 \times 10^{12}$  \\
Typical Beam Intensity (ppp)& $35 \times 10^{12}$ & $55 \times 10^{12}$ & $60 \times 10^{12}$ & $55 \times 10^{12}$ \\
Spill Length & $1.0$ sec & $1.6$ sec & $1.6$ sec & $1.6$ sec \\
Cycle Length & $3.8$ sec & $3.6$ sec & $3.6$ sec & $3.6$ sec \\
Duty Factor & $26\%$ & $44\%$ & $44\%$ & $44\%$ \\
Spill Structure Modulation & $50\%$ & $20\%$ & $<20\%$ & $20\%$ \\
Average Beam Current & $1.7 \mu A$ & $2.8 \mu A$ & $2.7 \mu A$ & $2.8 \mu A$ \\
Av. Availability/Best Week & $\sim 83\%$ & $82\%/93\%$ & $76\%/92\%$ & $72\%/79\%$ \\
\end{tabular}
\label{t: ags}
\end{table}

	Table.~\ref{t: ags} shows the recent performance of the
AGS~\cite{tomr} for slow extracted proton beam running. 
It is obvious that the intensity of the AGS has plateaued at about $6
\times 10^{13}$ protons per pulse.  Although this is a formidable
intensity, experiments have recently been discussed that could benefit
from even more.  In the last few years the work of T. Roser and others
has indicated that a further step can be made rather economically,
using a novel injection scheme~\cite{mblas} that requires very little
in the way of additional capital investment.  It is based on the use
of a broad-band RF cavity (``barrier bucket cavity'') to manipulate
the beam stored in the AGS.

\begin{figure}[htb]
\centerline{\epsfig{file=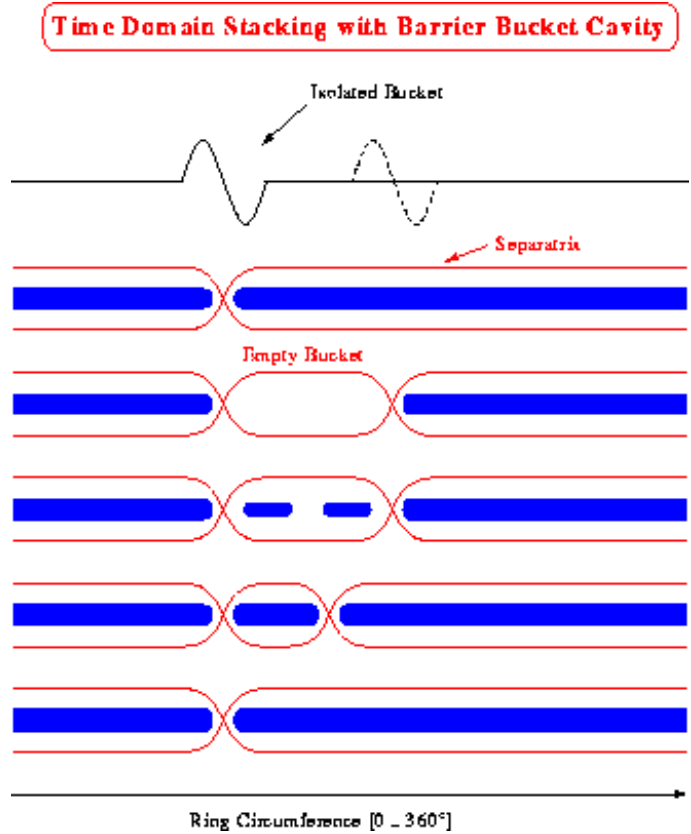,width=9cm,angle=0}}
\caption{\label{tds}
Time domain stacking with two barriers.}
\end{figure}

	  The AGS Booster accelerates 2 bunches 7.5 times a second,
injecting the bunches into the AGS where they are stacked in boxcar
fashion.  In the present injection scheme, the AGS accumulates four
successive Booster cycles (eight bunches) before accelerating.  During
the accumulation process, space charge effects begin to deplete the
stacked beam (the first bunches dwell 400 msec before acceleration
begins), and limit the total amount of beam that can be stored.  Since
it takes roughly 1 second to accelerate the beam in the AGS and
ramp the magnets back down, for fast extracted beam (FEB), the total
cycle is about 1+3/7.5 = 1.4 sec.  For slow extracted beam (SEB), the
cycle time is $\sim S + 1.4$, where $S$ is the length of the flattop.
Recently this has been 1.8 seconds with 1.6 seconds of useful spill;
with a few hundred $ms$ of overheads, the SEB cycle time is $\sim 3.6$
seconds.  At present, the Booster can supply $\sim 1.5 \times 10^{13}$
protons/cycle to the AGS, for a total of $6 \times 10^{13}$.  This
works out to $2.7 \> \mu$amp for SEB.

	In the proposed scheme, illustrated in Fig.~\ref{tds}, the beam
accumulated in the AGS is debunched.  After the first
Booster injection is debunched, an empty RF bucket is imposed on the beam,
creating a ``hole'' in the phase space.  A second empty bucket is
superimposed on the first, then slowly displaced, broadening the hole
in the beam.  A second Booster-load is then injected into this hole.
The two RF buckets are slowly moved towards one another until the
phase space density of the newly injected protons is equal to that of
the rest, after which one RF bucket is turned off.  The process is repeated
for each successive Booster injection.  Since the beam is debunched, about
twice as many protons can be accommodated under the space charge limit,
{\it i.e.} 8 Booster-loads, or 120 trillion protons (TP).  However, since
an extra 4/7.5 seconds is required to accelerate the additional
Booster-loads, one doesn't quite get twice the average current, and
the duty factor is reduced (see Table~\ref{t: upgr}).

\begin{table}[htb]
\centering
\caption{Possible AGS Intensity Upgrades}
\begin{tabular}{lll}
   &   \underbar{AGS only} & \underbar{AGS+Accum.}   \cr
\underbar{Slow Extracted Beam:} & & \cr
Duty Cycle (1s spill): & $33\%/28\%$ & $50\%$ \cr
Average Intensity: && \cr
4 Booster Cycl./AGS Cycle & 20 TP/s, 3.2$\mu$A & 30 TP/s, 4.8$\mu$A \cr
8 Booster Cycl./AGS Cycle & 34 TP/s, 5.4$\mu$A & 60 TP/s, 9.6$\mu$A \cr
 & & \cr
\underbar{Fast Extracted Beam:} & & \cr
Rep. Rate: & 1.7s/2.2s & 1.0 s  \cr
Average Intensity: && \cr
4 Booster Cycl./AGS Cycle & 35 TP/s, 5.6$\mu$A & 60 TP/s, 9.6$\mu$A  \cr
8 Booster Cycl./AGS Cycle & 54 TP/s, 8.6$\mu$A & 120 TP/s, 19.2$\mu$A  \cr
 & & \cr
\end{tabular}
\label{t: upgr}
\end{table}

\begin{figure}[ht]
\epsfysize=11cm
\centerline{\epsfbox{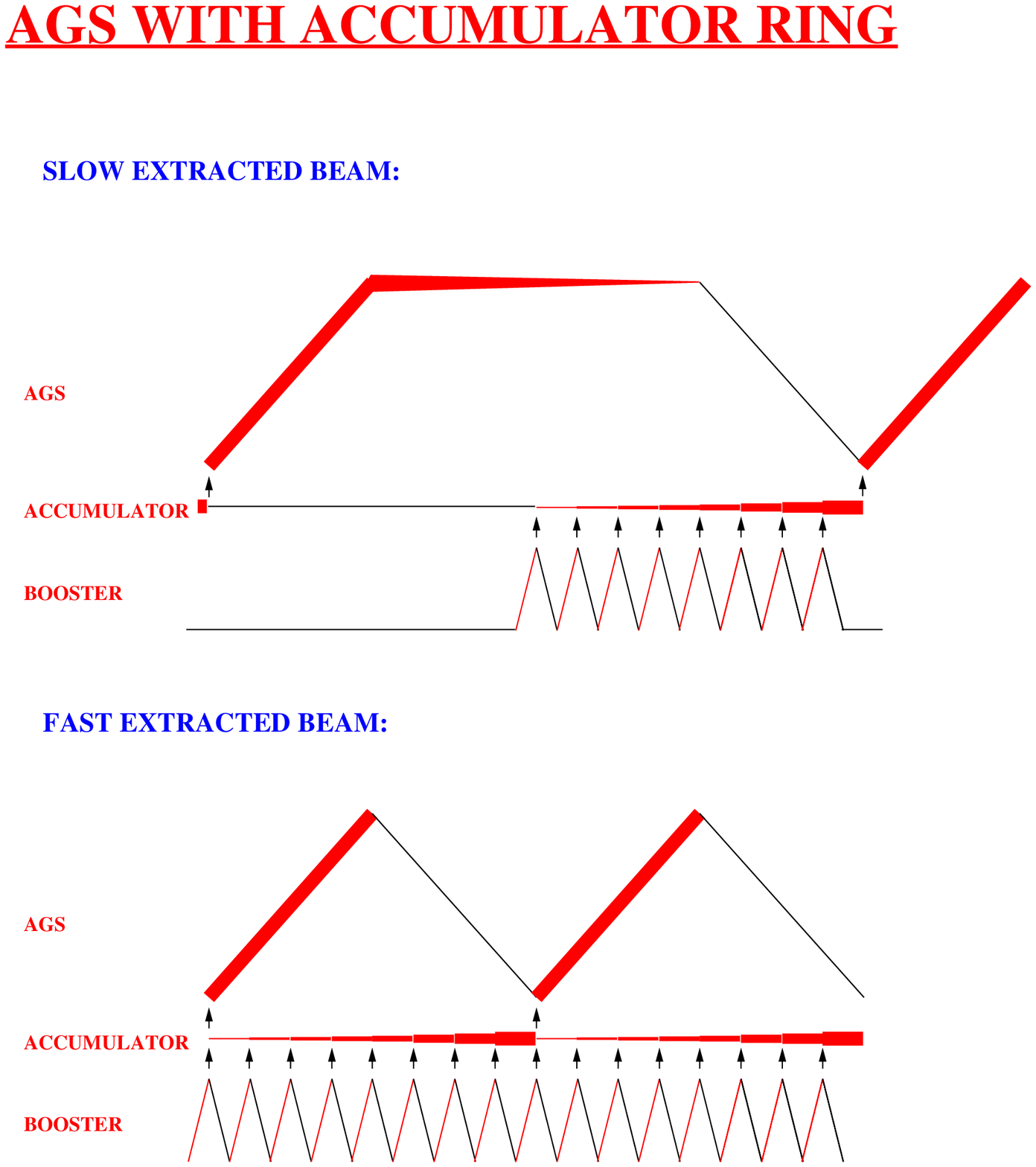}}
\caption{\label{acc}
Accumulator}
\end{figure}

	This scheme increases the average intensity of both fast and
slow beams by a factor $1.7$, at very modest cost.  A further improvement
can be made by eliminating the AGS accumulation time which will have grown
to $> 1$ sec, by adding a $1.5 \> GeV$ accumulator ring in the AGS
tunnel.  This is presently envisioned as a ring of 2.5 Kgauss
permanent magnets, installed above the present
ring.  Fig.~\ref{acc} shows the accumulation cycle for both fast and
slow extracted beam.  There would then be no contribution at all to
the AGS cycle time from the accumulation process.  For slow extracted
beam, this would yield a factor of up to 3 in average intensity with
respect to current performance, and improvement of the duty cycle to
$50\%$.  For fast extracted beam, the improvement would be even
greater, resulting in an intensity $20\%$ of that of the proposed
TRIUMF kaon factory.  It is possible that the performance of the AGS
could be increased even more than this, since the Booster intensity
already exceeds its original specification ($\sim 20 \> TP$ vs $15 \>
TP$).  Also, particularly for slow extracted beam, it might be possible
to accumulate more than eight Booster-loads.

	Table~\ref{t: upgr} gives the expected performance of the AGS
without and with accumulator, assuming four or eight Booster-loads of
15 TP each.  

	To summarize, for SEB running with the barrier bucket scheme
only, one can expect to reach 34 TP/sec, with a $28\%$ duty factor
(or, for example, 29 TP/sec with a $39\%$ duty factor).  Adding an
accumulator ring, one can reach 60 TP/sec with $50\%$ duty factor.
This is to be compared with recent running conditions of 17 TP/sec
with $44\%$ duty factor.

\section{The physics program}
\label{s: prog}

	The possible program for the AGS after 1999 was extensively
discussed at the AGS-2000 Workshop~\cite{ags2}.  The object was to
select a set of compelling experiments that either were unique to the
AGS, or were clearly best done there.  A number of promising
possibilities were identified.  Since the workshop, there has been a
further filtering of these possibilities.  Because almost all the most
interesting candidate experiments for this era will be covered by
separate talks at this workshop, I will be relatively brief in my
discussion of them.

\subsection{E940 (MECO)}

	Muon conversion in the field of a nucleus is a classic probe
of lepton flavor violation.  In this reaction, a stopped $\mu^-$ is
converted into an electron of energy $\approx m_{\mu}c^2$.  Since the
nucleus can remain in the ground state, coherence is possible, which
tends to make this more sensitive than other tests of lepton flavor
conservation such as $\mu \to e \gamma$.  A typical example is given
in Table~\ref{t: mue} which shows the reach of various types of
measurement for an interaction mediated by a generic horizontal gauge
boson\cite{cahara}.  One should note not only how well $\mu-e$
conversion stacks up against the competition, but also how large the
mass reach is in absolute terms, when one considers that the present
upper limit\cite{freip} is $ 7.8 \times 10^{-13}$.

\begin{table}[htb]
\centering
\caption{Relative sensitivity of various lepton-flavor-violating processes.}
\begin{tabular}{cc}
\cr
\underbar{~~~process~~~~} &\underbar{$M_H$ reach of a $10^{-12}$ experiment} \cr
 & \cr
$\mu^- N \to e^- N$& $300 \> TeV$ \cr
$\mu \to 3e $ & $~40\> TeV$ \cr
$\mu \to e \gamma$ & $~85\> TeV$ \cr
$K_L \to \mu e$ & $230\> TeV$ \cr
$K^+ \to \pi^+ \mu^+ e^-$ & $150\> TeV$ \cr
\end{tabular}
\label{t: mue}
\end{table}

\begin{figure}[ht]
\epsfysize=7cm
\centerline{\epsfbox{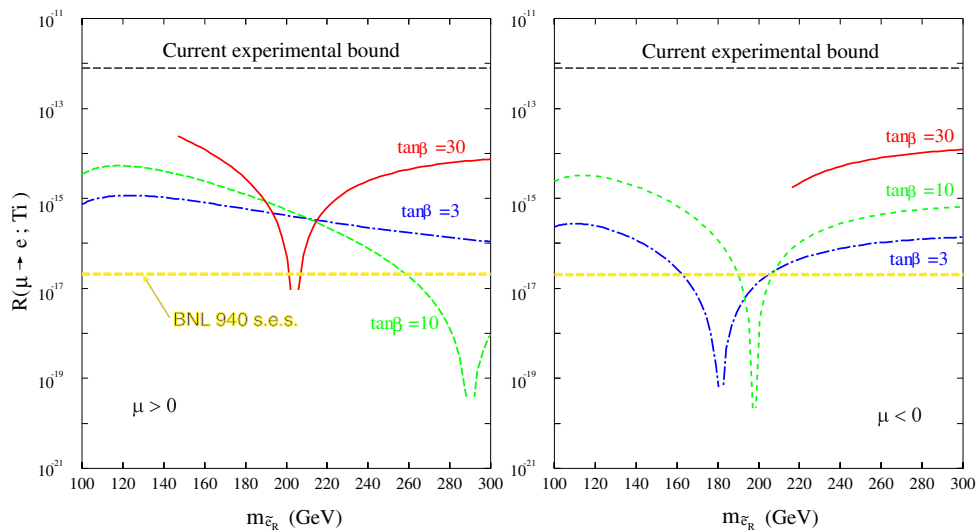}}
\caption{\label{themue}
Prediction of Ref.~\citen{hisano} for $\mu-e$ conversion for various values 
of the supersymmetric parameters compared with current upper limit and with 
projected single event sensitivity of AGS-940.}
\end{figure}

	The hierarchy of Table~\ref{t: mue} is model dependent, but
$\mu-e$ conversion generally comes out very well in this kind of
comparison, unless explicitly suppressed.  The coincidence of two
factors has greatly stimulated interest in this process.  Recent work
on flavor violation in supersymmetric GUTs\cite{muetheory,hisano} indicates
$\mu-e$ conversion at levels two or three orders of magnitude below the
current upper limit would be very natural.  This is illustrated in
Fig.~\ref{themue}.  Secondly, proponents of 
muon colliders have emphasized the very large production of low energy
pions (and therefore muons) by medium energy proton beams striking heavy
targets.  Since the signature for $\mu-e$ conversion, the appearance
of an electron of energy $\approx~m_{\mu}c^2$ out of a stopped muon
beam, does not require a coincidence experiment, an extremely interesting
opportunity arises.

\begin{figure}[ht]
\epsfysize=10cm
\centerline{\epsfbox{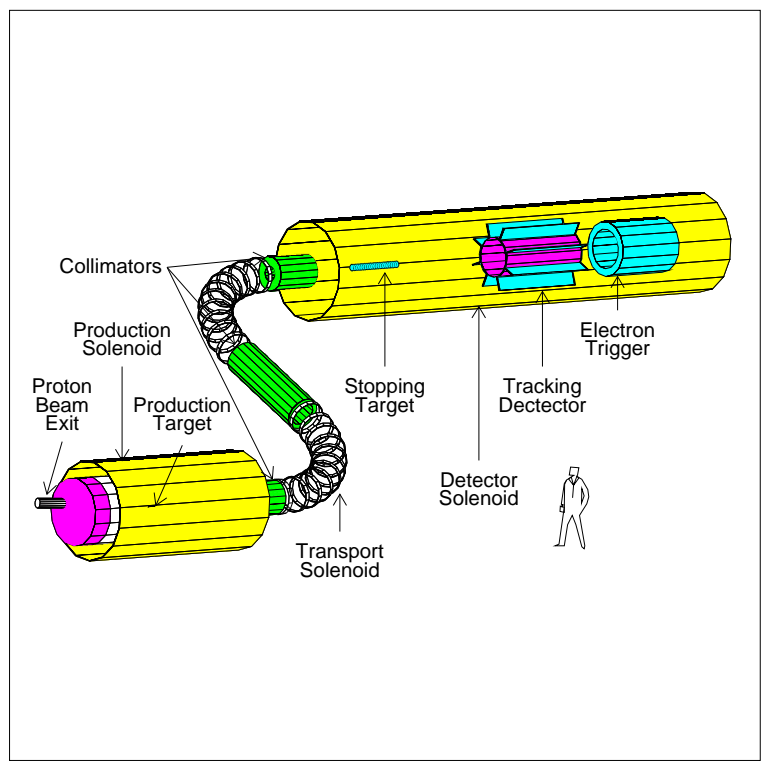}}
\caption{\label{medet}
Schematic of proposed AGS-940 detector.}
\end{figure}

	Bill Molzon and his collaborators have seized the moment to
propose a new experiment~\cite{molzon}, based to a large extent on the
MELC proposal\cite{melc} at the Moscow Meson Factory.  Fig.~\ref{medet}
is a schematic of their proposed detector.  It is composed of three
large solenoidal field regions.  A proton beam of $4 \times
10^{13}$/pulse enters the production solenoid from the right and
impinges on a heavy production target.  The time structure of the beam
is a pulse of a few nanoseconds every microsecond.  This is done so
that one can ``wait out'' $\pi^-$'s that might otherwise make background
through radiative pion capture.  One can gate off the detector for the
first few hundred $ns$ until all pions decay.  To accomplish this, there
must be fewer than $1$ in $10^{9}$ ``breakthrough'' protons between
pulses, a very difficult task for the AGS.  Most probably the pulsed
structure have to be supplemented with some other technique, such as a
pulsed kicker.  The large number of low energy pions created in the
pulse are efficiently collected by the graded solenoidal field in the
production solenoid.  Of course the pions are decaying constantly to
muons so that during the collection process one has an ensemble of
pions and muons, that eventually becomes all muons.  Muons are
transported and separated by sign by the transport solenoid, and
negative muons delivered to the detector solenoid.  The aim is to
capture on the order of 1 negative muon per 100 incident protons.  The
muons are captured on a series of thin Al stopping targets.  Once
again a graded solenoidal field is used to collect the resulting
electrons with maximum efficiency.  The detector is designed to be
blind to electrons from ordinary muon decay which are confined by
the field to radii inside the region of detectors.  A straw chamber
or scintillating fiber tracking array followed by an electron
trigger calorimeter are proposed.

	Other possible backgrounds are radiative muon capture, muon
decay in flight, scattered beam electrons, and cosmic rays.  Recently
it was pointed out that antiprotons could possibly
constitute a significant source of background.

\subsection{$K^+ \to \pi^+ \nu\bar\nu$}

	This decay mode is extremely interesting because it first arises
at the one electroweak loop level in the Standard Model.  As a consequence
it is suppressed by some nine orders of magnitude with respect to
the kinematically identical $Ke3$ decay.  This makes for a very sensitive
probe for new phenomena, of which many have been proposed\cite{BSM}.
In the SM, the branching ratio is given by 

\beqn
B(K^+\rightarrow\pi^+\nu\bar\nu) &=&
{{\alpha^2 B(K^+\rightarrow \pi^0 e^+ \nu)}\over{V_{us}^2 2\pi^2sin^4\theta_W}}
\sum_{\ell} |V^*_{cs}V_{cd}X^{\ell}_{NL}+V^*_{ts}  V_{td} X(x_t)|^2
\label{pnnbr}
\eeqn
\noindent
where $X(x_t)$ is a kinematic function of the top quark mass,
and $X^{\ell}_{NL}$ is a QCD-corrected kinematic function of the
charm quark mass.  The hadronic matrix element, so problematical in
other weak decays, is determined to $\cal O$$(1\%)$ from the
well-measured $Ke3$ decay rate\cite{2mar}. In the current experimental
and theoretical situation, the most interesting potential is that of
determining $|V_{td}|$ from a measurement of this branching ratio.
After next-to-leading-logarithmic order QCD corrections, the intrinsic
theoretical uncertainty in $|V_{td}|$, given
$B(K^+\rightarrow\pi^+\nu\bar{\nu})$, is $<5\%$~\cite{bandb}, driven
mainly by uncertainty in the charm contribution.  Of course since
$|V_{td}|$ and other input parameters are not yet tied down, the
present {\it prediction} of the branching ratio is much more uncertain.
Imposing constraints
from $\bar B-B$ mixing and from assuming the SM origin of CP-violation
in the $K$ system, the current estimate of this branching ratio is
$(0.6 - 1.5) \times 10^{-10}$\cite{BF}.

\begin{figure}[ht]
\epsfysize=8cm
\centerline{\epsfbox{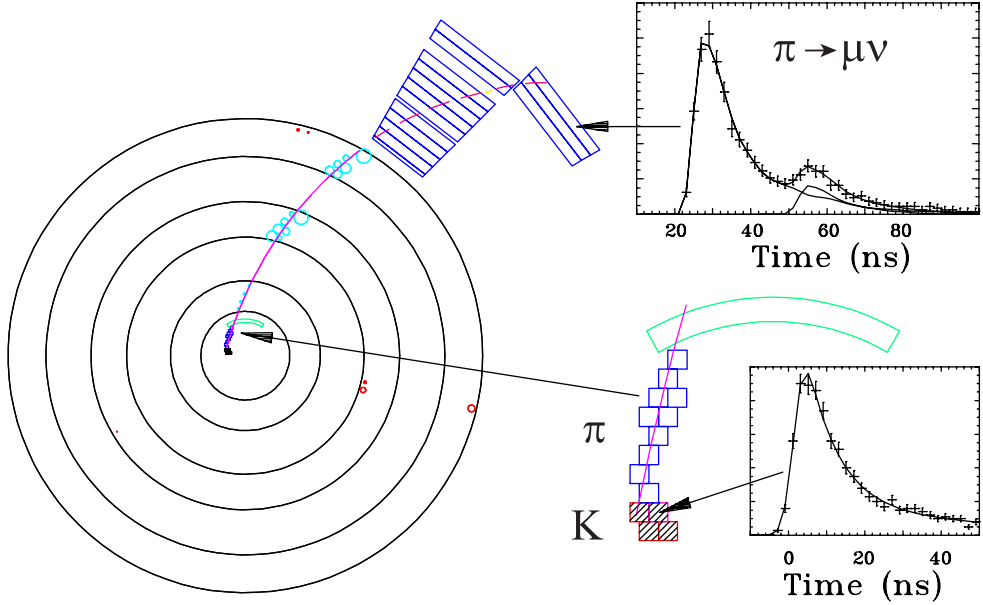}}
\caption{\label{event}
E787 candidate for $K^+ \to \pi^+ \nu\bar\nu$.}
\end{figure}

	As everyone here knows, AGS-787 has announced the observation
of a strong candidate for $K^+ \to \pi^+ \nu\bar\nu$~\cite{adler} (see
Fig~\ref{event}).  That this milestone could be reached with very low
background, opens the door to exploiting kaon flavor changing neutral
current reactions to make precise measurements of CKM quantities.  Of
course, we still aren't $100\%$ certain that the Standard Model
applies!  The near-term task of AGS-787 is establish whether in fact
it does.  The branching ratio implied by the observed event, $(4.2
{+9.7 \atop -3.5}) \times 10^{-10}$, is consistent within statistics
with the Standard Model (SM) estimate, although it is $3-4$ times
higher.  If the central SM prediction is correct, then, as I-Hung
Chiang~\cite{ihc} will discuss later in this workshop, E787 is
unlikely to get more than a handful of events by the end of 1999.
Since $K^+ \to \pi^+ \nu\bar\nu$ offers probably the theoretically
cleanest method of determining $|V_{td}|$, it would be a shame not be
able to exploit it fully.  Roughly speaking, the relative error on
$|V_{td}|$ is about $\frac{2}{3}$ that on $B(K^+ \to \pi^+
\nu\bar\nu)$.  Thus, a signal of 5 events would determine $|V_{td}|$
to about $\pm 30\%$.  If one could get an additional factor $5$ in
statistics via an AGS upgrade, this would go to $< 15\%$, which would
be sensational!  Since John Macdonald~\cite{jam} will discuss the
future plans of this collaboration to pursue $K^+ \to \pi^+
\nu\bar\nu$ in the AGS-2000 era, I will be relatively brief.

\begin{figure}[ht]
%\epsfysize=10cm
%\centerline{\epsfbox{manny_new.eps}}
\centerline{\epsfig{file=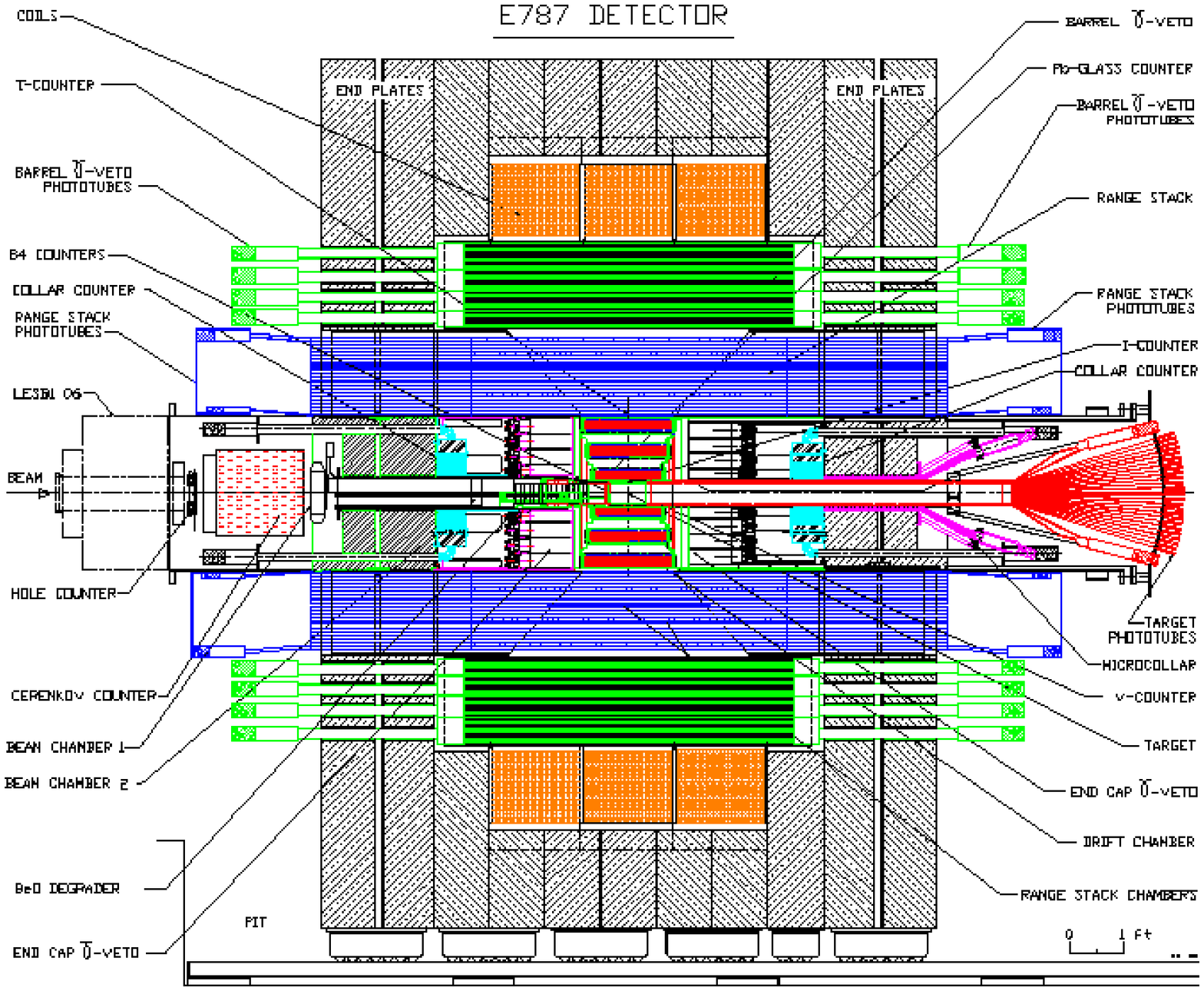,width=10cm,angle=90}}
\caption{\label{d787}
Schematic of AGS-787 detector.}
\end{figure}

Fig.~\ref{d787} shows the E787 detector.  Major upgrades are not thought
to be necessary in order to make substantial further progress in the
study of \kpinn.

	The plan for future running includes lengthening the present
AGS spill so that the experiment can accumulate more sensitivity per
hour without increasing its instantaneous rate capability.  If a total
of $60 \>TP$ could be devoted to E787, a factor $\sim 1.7$ increase
would be possible.  This is what allows the experiment to project a
single event sensitivity in the $2-4 \times 10^{-11}$ range in the
near term.  Simply turning up the wick by another factor five would
not be easy.  Trigger and random veto dead-times, not to speak of
off-line background, would become very hard to fight at that level.
Instead, to exploit an increase of available protons one might
consider reducing the incident $K^+$ beam
momentum from $700 \> MeV/c$ to perhaps $550 \> MeV/c$.  This would
allow a substantially greater yield of stopping $K^+$ per incident
$K^+$.  Right now this ratio is only about $\frac{1}{4}$.  It might be
possible to make it as high as $\frac{1}{2.5}$.  In this case, for about twice
the number of protons on the production target one would have the same
number of $K^+$ incident on the degrader, but $\sim 1.6$ times as many
would be stopped.  We are confident of being able to sustain this rate
since we have found through several changes of the beam momentum 
that instantaneous rates in the
detector are roughly proportional to the rate of $K^+$ which
interact in the degrader.  If the available proton flux were not
sufficient to do this {\it and} to lengthen the spill, a shorter, ($14.7 \> m$
vs the present $19.4 \>m$), higher-acceptance beamline could be built,
which would provide a factor two in $K^+$ per proton.  Another
measure would be to exploit the phase space for $K^+ \to \pi^+
\nu\bar\nu$ with $p_{\pi} < 205 \> MeV/c$.  Presently, we use only the
$20\%$ of phase space with $p_{\pi} > 205 \> MeV/c$, since there are
no common $K^+$ decay modes that produce $\pi^+$ with momenta above
this value.  The E787 detector already samples quite a large region
below $205 \> MeV/c$.  Potentially, the acceptance of the experiment
could be more than doubled.  The main barrier is \kpitwo (``$K \pi
2$'') decays in which the $\pi^+$ loses energy undetectably in the
stopping target, while the $\pi^0$ eludes the photon veto.  The
stopping target instrumentation has been much improved in recent years
and we are hoping (but not promising!) to be able to use this region
in our current data.  For the future, a scheduled upgrade to the
photon veto should help clean up this region, and other expedients are
being discussed.  The final factor would come in running time.
RHIC is scheduled to run 37 weeks/year.  If we could use the AGS
for 35 of these weeks, we could double the effective running time of
our best year so far.

	Thus a path is open to improve the precision on $K^+ \to \pi^+
\nu\bar\nu$ to $< \pm 20 \%$, and that on $|V_{td}|$ to $< 15 \%$,
without major upgrades to E787, as long as enough protons could be made
available.

\subsection{$K_L \to \pi^0 \nu\bar\nu$}

The branching ratio for $K_L \to \pi^0 \nu\bar\nu$ is given by
Eq.\ref{pnnbr} with the moduli of the CKM terms replaced by their
imaginary parts.  This essentially removes the charm term, and with
it, almost all the residual theoretical uncertainty. The uncertainty
in the branching ratio given the input parameters is reduced from
$\sim 7\%$ in the charged to $\sim 1\%$ in the neutral case.  In terms
of the Wolfenstein representation \cite{Wolf}, $V_{td} = \lambda^3 A
(1 - \rho - i \eta) $, where $\eta$ characterizes CP violation in weak
decays.  $B(K^+ \to \pi^+ \nu\bar\nu)$ is sensitive to both $\rho$ and
$\eta$, while $B(K_L \to \pi^0 \nu\bar\nu)$ is sensitive only to
$\eta$. The contribution of indirect CP-violation to $K_L
\rightarrow\pi^0\nu\bar{\nu}$ is tiny\cite{ll} as are the
long-distance contributions\cite{hal}, so that a measurement of the
neutral branching ratio will yield an accurate determination of
$\eta$, given 3-generation unitarity and knowledge of $|V_{cb}|$
(since the rate is actually proportional to $[Im({V_{ts}^* V_{td}})]^2$).  
The current estimate is $B(K_L \rightarrow\pi^0\nu\bar{\nu}) = 
(3 \pm 2) \times 10^{-11}$, where once again the extent of the range is 
given by uncertainties in the input parameters.

\begin{figure}[htb]
\center\epsfig{file=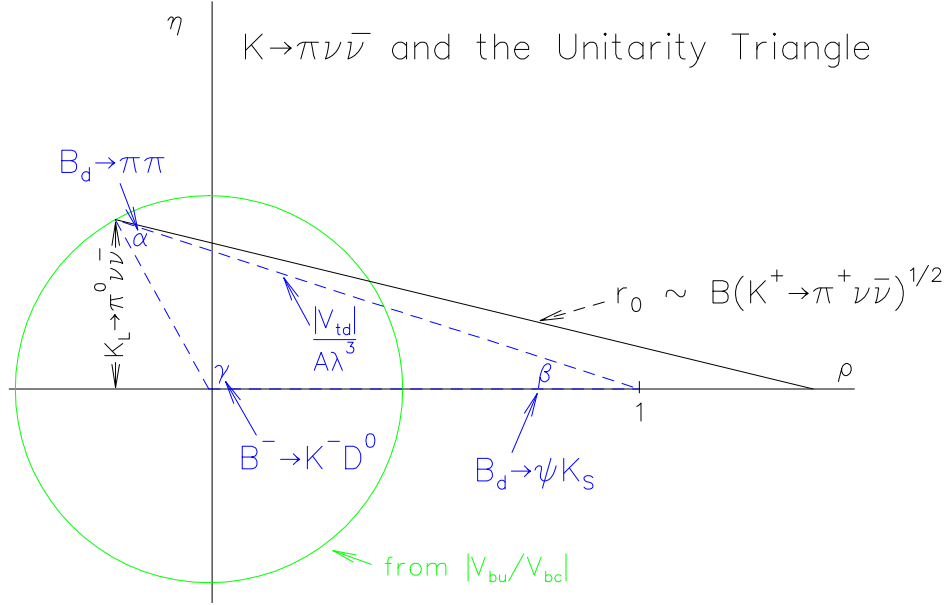,width=80mm,angle=90}
\caption{Diagram of the contribution of the charged and neutral
FCNC kaon decay $K\rightarrow\pi\nu\bar\nu$.}
\label{UT}
\end{figure}

If one can add to this a measurement of
$B(K^+\rightarrow\pi^+\nu\bar{\nu})$, which as discussed above, is
dominated by a term proportional to $|V_{ts}^* V_{td}|^2$, one can
determine the unitary angle $\beta$, independent of measurements in
the $B$ system\cite{bbang,BuchBur}.  Taking a ratio of the neutral to
charged branching ratio removes any uncertainty due to $V_{ts}$.  Figure
\ref{UT} illustrates the relationship between the unitarity triangle
and the two kaon FCNC rates.

\begin{figure}[htb]
\center\epsfig{file=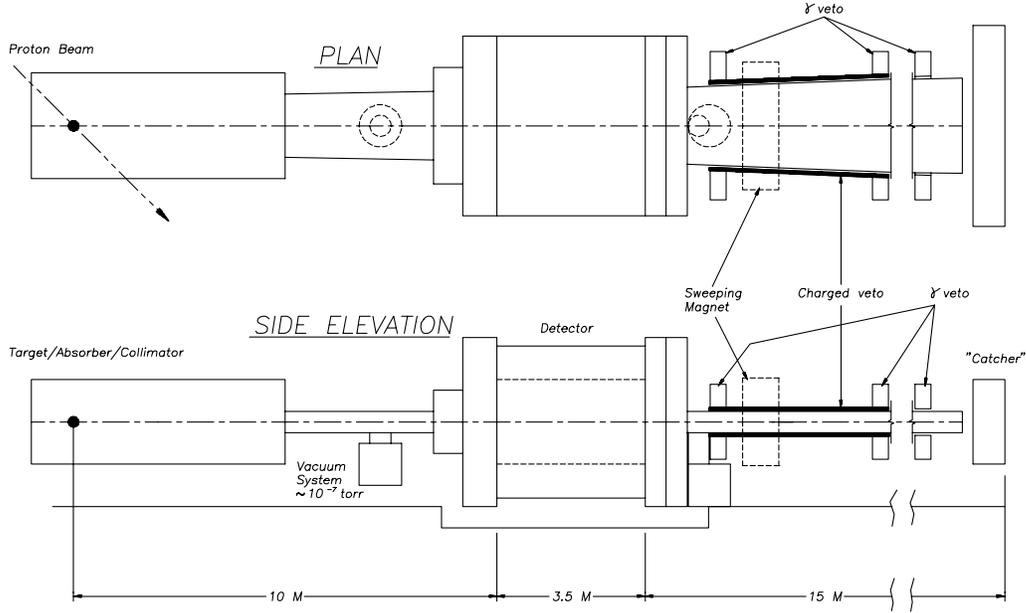,
width=135mm,angle=0}
\caption{Schematic of the proposed 926 detector.}
\label{d926}
\end{figure}

	An experiment to measure $B(K_L \to \pi^0 \nu\bar\nu)$ that
exploits the strengths of the AGS was approved in 1996\cite{p926}.  A
schematic of the proposed detector is shown in Fig.~\ref{d926}  Using
$0.5 \times 10^{14}$ protons per AGS acceleration cycle, it is
estimated that in 80 weeks of running time, about $70$
$K^0\rightarrow\pi^0\nu\bar{\nu}$ events could be collected with a
background of $\approx 10\%$.  This would yield a precision on $\eta$
of $< 10\%$ (if CKM $A$ were perfectly known).  The techniques
proposed are as follows.  (1) The neutral beam will be extracted at an
extremely large angle ($\sim 45^o$) to soften both the neutron and
kaon momentum spectra.  This minimizes the flux of neutrons capable of
producing background by interacting with vacuum windows or residual gas.
To further suppress background from the latter, a vacuum of $10^{-7}$
Torr must be maintained throughout the beam region.  (2) The beam will
be highly asymmetric ($4 \> mr \times 125 \> mr$) to facilitate
shielding and afford an extra kinematic constraint.  (3) The AGS
proton beam will be bunched on extraction with rms $\leq 200 \> ps$ every
$\sim 40 \> ns$.  The narrow bunch width allows the use of flight time
measurement to determine the neutral kaon momentum to a few percent.
The soft kaon spectrum ($ \bar p \sim 750 \> MeV/c$) is crucial in
allowing this technique to work in a short ($10 \> m$) beamline.  In
addition to the momentum determination, due to the microbunching the
massless and other fast debris from the target interaction will arrive
at the detector before the kaons of interest, and so can be
distinguished from the latters' decay products.  (4) Active shower
pre-converters will measure the direction of $\pi^0$ photons emanating
from the $K_L\rightarrow\pi^0\nu\bar\nu$ decay.  (5) In conjunction
with a high resolution scintillating fiber calorimeter based on the
KLOE design~\cite{kloe}, this allows one to fully reconstruct the
$\pi^0$, independent of any assumptions about the beam.  Combining the
$\pi^0$ with the beam timing information, one can transform into the
$K_L$ center of mass.  (6) Hermetic photon vetoing is required.  A
very conservative extrapolation from photon vetoing performance
measured in E787 indicates that an average rejection of $10^4:1$ per
$\gamma$ can be achieved.  The largest expected background to $K_L
\rightarrow\pi^0\nu\bar{\nu}$ is the 300-million times more frequent
$K_L \to \pi^0 \pi^0$ decay ($K_{\pi 2}$) when two of the four final
state photons are missed.  When the two missed photons come from the
decay of the same $\pi^0$ (``even'' case), the two detected photons
will reconstruct properly to a $\pi^0$. The energy of this
reconstructed $\pi^0$, in the center of mass system of the $K_L$, will
equal $249 \> MeV/c $ (modulo the resolution).  When the two detected
photons come from different $\pi^0$'s (``odd'' case), they will not
generally have an effective mass $\approx m_{\pi}$.  It's also an
important advantage of low energy kinematics that unvetoed $K_{\pi2}$
events tend to have rather small values of missing energy and missing
mass compared to signal events.  As a result, $K_{\pi 2}$ background
can be suppressed to $\leq 10\%$ of the signal.

Other possible backgrounds are $K_L \rightarrow \gamma\gamma$, 
$K_L \rightarrow \pi^- e^+ \nu$, with both charged daughters interacting, 
$\Lambda \rightarrow \pi^0 n$, and accidental $\pi^0$'s. These
backgrounds have been calculated to to be much smaller than that
from $K \pi2$.

	Later in this workshop, Yury Kudenko will talk on recent progress
on E926\cite{p926} including the results of beam and prototype tests.

\subsection{$T$-violating $\mu^+$ polarization in $K^+ \mu3$ decay}

In his talk on KEK-246 at this workshop, Masa Aoki~\cite{kek246}
discussed the motivation for searching for a $T$-violating polarization
in $K^+ \to \pi^0 \mu^+ \nu$.  Briefly, the need for CP-violation
in addition to that given by the SM in order to explain the
observed baryon asymmetry of the universe\cite{basu} motivates investigating
low-energy `windows' where such effects are cleanly identifiable.
The CKM matrix gives virtually no T-violating polarization in 
$K^+ \to \pi^0 \mu^+ \nu$, opening such a window.  Moreover a number
of popular attempts to go beyond the Standard Model predict a finite
polarization at a level that is experimentally accessible\cite{km3th}.

\begin{figure}[htb]
\center\epsfig{file=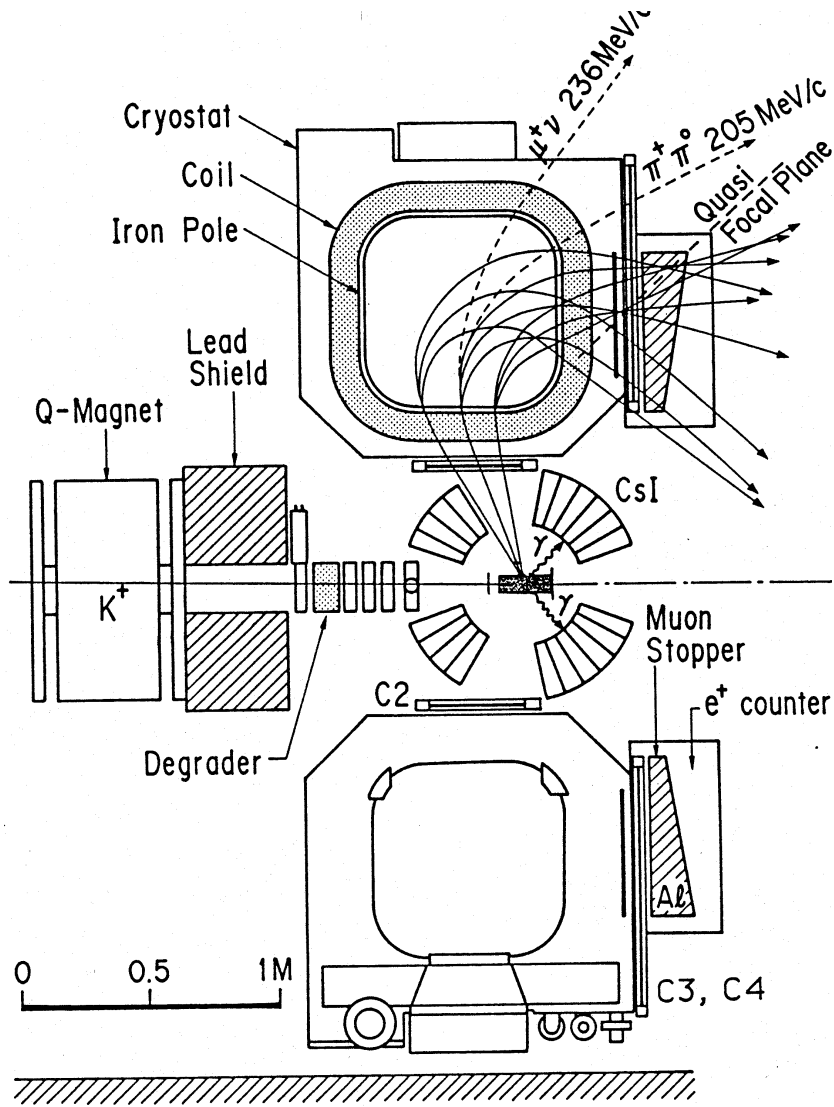,
width=110mm,angle=0}
\caption{Schematic of KEK-246 experiment.}
\label{d246}
\end{figure}

	If $T$ invariance is not violated, the $f_+(q^2)$ and 
$f_-(q^2)$ form factors which multiply the $(p_K  + p_{\pi})$
and  $(p_K  - p_{\pi})$ terms in the $K \mu 3$ amplitude
are relatively real.  Therefore $T$ violation is characterized
by the size of the imaginary part of their ratio $\Im{\xi} \equiv
\Im{(f_-/f_+)}$.  This quantity is in turn approximately proportional
to the component of polarization transverse to the $K \mu 3$ decay
plane, $\wp_T = (0.2-0.3) \, \Im{\xi}$ depending on the phase space
sampled.

	KEK-246 is very familiar to this audience.  Fig~\ref{d246} is
a schematic of the detector.  A $660 \> MeV/c$ separated $K^+$ beam is
slowed in a degrader and stops in a highly segmented scintillating
fiber target.  The trigger is designed to accept $K^+ \to \mu^+ \pi^0
\nu$ decays at rest.  Photons from the $\pi^0$ are detected in a
768-element CsI(Tl) counter array.  The muons are tracked and momentum
analyzed in a spectrometer built around a superconducting toroidal
magnet with 12-fold symmetry.  The muons are slowed in wedge-shaped Cu
degraders and stop in sets of Al plates at the exit of each of the
magnet's 12 gaps.  Counter arrays surrounding the Al stoppers complete
the polarimeters.  The polarimeters are used to search for tracks which
enter (but do not exit) at $K^+$ decay time, followed by delayed tracks
exiting transversely.  Tracks exiting left and right with respect to
the entering muon direction are compared.  The fringe field is aligned
with the $\wp_T$ direction, pinning the transverse polarization while
precessing the allowed longitudinal polarization.  The 12-fold
symmetry of the detector cancels or at least reduces many possible
systematic errors.  Summing the results from the 12 polarimeters, one
searches for a difference between clockwise and counterclockwise-going
decays.  Great care has been taken on alignment and control of
magnetic fields.

In addition to the 12-fold symmetry of the muon system, because of the
stopping geometry, one also has a forward-backward symmetry for the
$\pi^0$ direction that can be exploited to control systematic errors.
Unfortunately the limits on running time and available proton flux at
KEK make it impossible for this experiment to reach its full
potential, although it is expected to reach $\sigma_{\wp_T} = \pm
0.0013$.  Since it is believed that the apparatus is capable of much
more, the experimenters have submitted a proposal for transporting the
experiment to the AGS\cite{p936}, where a result some four times more
sensitive is possible.  At LESB3 they could get $5 \times
10^6~K^+$/pulse with a $\pi/K$ ratio of $<0.3$ which should be
compared to $3 \times 10^5~K^+$/pulse and $\pi/K \sim 7:1$ at KEK.
This means that most singles rates in the experiment will be
comparable at BNL and at KEK, but there will be $\sim 15$ times more
usable $K^+$ decays/pulse.  Rates proportional mainly to $K$ decays
will of course be much higher than at KEK, so that, for example, the
CsI shaping time will have to be reduced, more instrumentation will
need to be deployed for the MWPCs, and the trigger will have to be
upgraded.  All in all, however, the necessary modifications to the
apparatus are quite modest.  The experimenters aim to collect some 
$3 \times 10^8~K \mu 3$ events in $3000$ hours of running, to
achieve a statistical sensitivity of $\pm 0.00035$ on $\wp_T$.

\begin{figure}[htb]
\center\epsfig{file=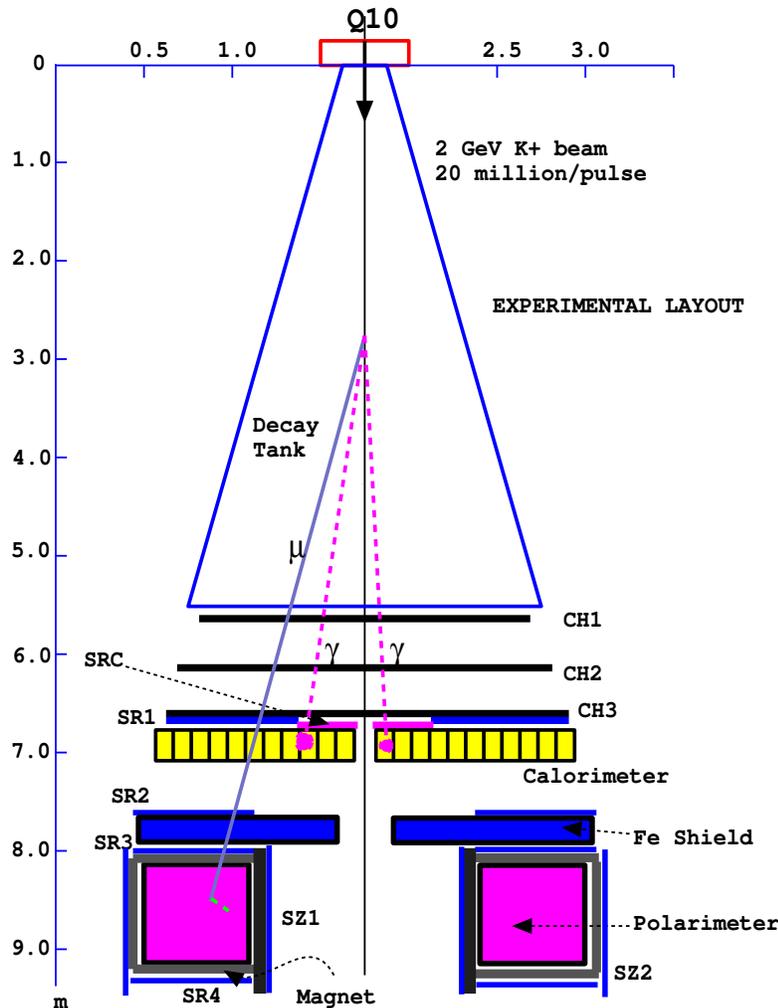,
width=135mm,angle=-90}
\caption{Schematic of the proposed 923 detector with a $K \mu3$ event
superimposed.}
\label{d923}
\end{figure}

KEK-246 represents a new technique in the study of T-violating $\mu^+$
polarization in $K\mu3$.  It looks promising, but it has not quite
proved itself yet.  A second approach~\cite{milind} being advocated is
to instead push the technique of the most recent previous experiment
of this type~\cite{lastpol} (also carried out at the AGS).
Fig.~\ref{d923} shows the layout of the proposed experiment.  The
biggest single improvement with respect to Ref.~\citen{lastpol} is the
use of a $2 \> GeV/c$ separated beam.  Other improvements include a
much larger acceptance, a more nearly complete reconstruction of the
decays, a more finely segmented polarimeter, and the use of graphite,
instead of aluminum as the polarimeter muon-absorbing material.  A beam of
approximately $20$ million $2 \> GeV/c~K^+$'s/AGS pulse is incident on a
decay tank in which about $5$ million decay.  Photons from the
$\pi^0$ decay are detected in a Pb-scintillator calorimeter and
$\mu^+$'s penetrate the calorimeter and are tracked into the
polarimeter where they stop.  These muons eventually decay and their
daughter electrons are tracked through at least two segments of the
cylindrically symmetric polarimeter.  The principle here is similar to
that of KEK-246 -- once again one is looking for differences in the
rates of clockwise-going and counter-clockwise-going muon decays.  In
this case, there are 96 segments as compared to 12 in KEK-246.  To
properly align the the decay plane with the detector, $K^+$ decays
where the $\pi^0$ emerges along the beam and the $\mu^+$ roughly
perpendicular to it in the $K^+$ center of mass are selected by the
trigger.  There is no spectrometer magnet, but a $70$ G
solenoidal field is imposed on the polarimeter to precess the muons.
The polarity of this field is reversed every AGS pulse.  This
technique is very effective in controlling systematic errors.  The
analyzing power of the polarimeter is calculated to be $\cal
O$$(30\%)$ which is a large improvement over that of
Ref.~\citen{lastpol}.  The expected statistical sensitivity of the
experiment is $\sigma_{\wp_T} = 0.00013$ in about 2000 hours of
running.  This corresponds to an uncertainty of roughly $7 \times
10^{-4} $ in $\Im{\xi}$.  Large efforts have gone into studying and
minimizing possible systematic errors.

	Table~\ref{t: kmu3} compares KEK-246, AGS-936, and AGS-923.

\begin{table}[htb]
\centering
\caption{T-violating muon polarization in $K \mu 3$}
\begin{tabular}{llll}
 & \underbar{KEK-246}   &   \underbar{AGS-936} & \underbar{AGS-923} \\
$\sigma_{\wp_T}$ & 0.0013 & 0.00035 & 0.00013 \\
beam & stopping & stopping & 2 GeV/c sep. \\
recons. evts & yes &yes &partial \\
$\mu^+$ stopper & Al & Al & graphite \\
$e$-detector & scintillator & scintillator &Al drift chamber or scint. \\
precess $\wp_T$? & no & no & yes \\
precess both allowed $\wp$? & yes & yes & no \\
fwd/bckwd symmetry? & yes & yes & no \\
limited by & statistics & statistics & systematics \\
comments & beam $\frac{\pi}{K} = \frac{7}{1}$ & CsI(Tl) rates& 
rate sensitive \\
\end{tabular}
\label{t: kmu3}
\end{table}

	Both AGS-923 and AGS-926 will also attempt to measure the
T-violating muon polarization in $K^+ \to \mu^+ \nu \gamma$

\subsection{Children of $(g-2)$}

	There has been a long-standing project at the AGS to improve
the measurement of the muon anomalous magnetic moment (g-2) to $0.35$
parts per million\cite{gm2}.  This represents an improvement by a
factor 20 over that of the previous (CERN) experiment\cite{cern}.
AGS-821, which took its first data in 1997, is built around a 7.11m
radius superferric storage ring in the form of a continuous ``C''
magnet with the open side toward the center of the ring.  The central
field is $1.45 \> T$, and it is designed to store muons at the
``magic'' momentum of $3.094 \> GeV/c$.  Lead-scintillating fiber
calorimeters are situated symmetrically at 24 sites around the inside
of the ring to detect decay electrons.

	Aside from the primary goal of the experiment, it is intended to
test CPT by measuring the difference between the $(g-2)$ of positive and
negative muons.  It is also intended to improve the current limit on 
the muon electric dipole moment (EDM) by an order of magnitude.  Going
beyond this, there have been two serious suggestions for reconfiguring
the apparatus to serve new experiments.

\subsubsection{Direct measurement of $m_{\nu_{\mu}}$}

	There's interest in using the $(g-2)$ ring as a spectrometer
to improve the sensitivity to the mass of the muon neutrino by more than
order of magnitude\cite{prisca}.  The precision of the most recent 
previous searches\cite{rmmunu} was limited by the uncertainty in the
pion mass to $m_{\nu_{\mu}}^2 < 0.16 \> MeV$.  In an in-flight measurement
at $p_{\pi} \sim 3 \>GeV/c$, this source of uncertainty becomes negligible.

\begin{figure}[htb]
\center\epsfig{file=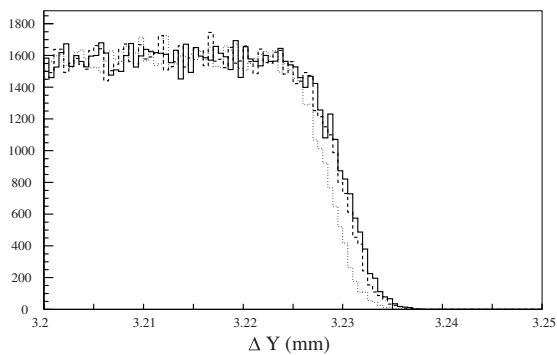,
width =80mm,angle=0}
\caption{Monte Carlo simulation for the proposal to measure the 
$\nu_{\mu}$ mass in the $(g-2)$ ring.  The end point region of the 
position distribution of muons at the microstrip detector is shown.
Abscissa is the distance from the parent pion position in $mm$.  Solid line
is distribution for the case of $m_{\nu_{\mu}} = 0$, dashed line for the
case of $m_{\nu_{\mu}} = 20 \> keV$ and dotted line for the case of
$m_{\nu_{\mu}} = 40 \> keV$.}
\label{endpt}
\end{figure}

Such a measurement requires position information on an event-by-event 
basis, and thus requires some changes to the septa and the AGS tune
in order to bring a slow-extracted pion beam to the
$(g-2)$ ring. The pions will be injected into the ring, much as was done for
the 1997 g-2 run. They are then ``kicked" into the proper orbit by
degrading their energy in 9 cm of beryllium sandwiched between two 
microstrip detectors.  One has to measure
a tiny shift in transverse position of the decay muon after one turn in the
ring as it passes its parent pion's entering point to the detector sandwich.  
Muons of the maximum momentum will
deviate by about 3mm from the pion position if $m_{\nu_{\mu}} = 0$.
This will be the endpoint of the spectrum of muon positions, and the
game is to determine this endpoint to extremely good precision
(see Fig.~\ref{endpt}).  If a
resolution of $1.4 \mu m$ can be attained, one can reach a $2 \>
\sigma$ sensitivity of $8 \> keV$ for $m_{\nu_{\mu}}$.  This assumes
detection of a total of $2 \times 10^{12}$ pion decays, which is
estimated to require about 800 hours of beam time.

\subsubsection{A measurement of the muon EDM to $10^{-24}\;e\>cm$}

	There is a letter of intent\cite{edm} to measure the muon
electric dipole moment with a precision of $10^{-24}\;e\>cm$, which
represents a factor $10^6$ improvement over the current state of the
art.  At this level, the measurement becomes interesting from the
point of view of supersymmetry\cite{edm_th} and competitive in
constraining power with current limits on the electron and neutron
EDMs.  

\begin{figure}[htb]
\center\epsfig{file=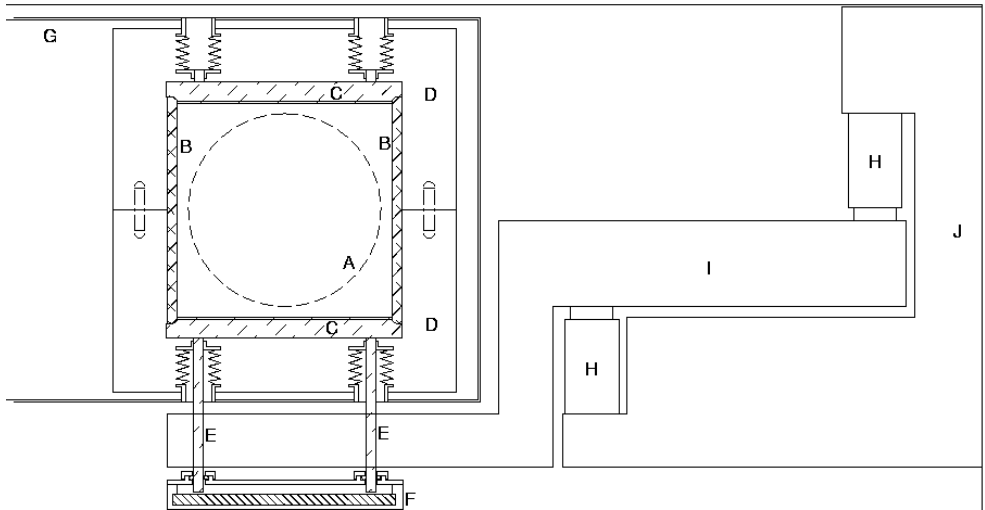,
width=67.5mm,angle=90}
\caption{Scheme of electrode alignment showing: (A) muon storage region;
(B) electrodes; (C) conductive glass plates with insulating support webbing;
(D) ceramic alignment yokes; (E) glass reference rods; (F) hydrostatic 
reference surface; (G) vacuum chamber; (H) piezoelectric driver; (I) level;
and (J) support beam.}
\label{alig}
\end{figure}

	To make a measurement at this level, it is proposed to reduce
the $(g-2)$ ring momentum from the magic value to $\sim 500 \> MeV/c$
at which point it is practical to create a radial electric field large
enough to cancel the $(g-2)$ precession.  The EDM precession then
operates without competition.  As in the case of $(g-2)$, the muon's
parity-violating decay is used to measure its spin direction.  One
looks for an asymmetry between upward and downward-going electrons as
a function of time.  This can be done to some level with the present
electron detectors, but to reduce systematic errors associated with
the controlling the average vertical position of the muons, detectors
will be placed on the top and bottom of the vacuum chamber.  These
will also be optimized for the lower electron energies.  Great efforts
must be made to control unwanted components of the electric field.
The precision needed on mechanical alignment is rather challenging.
Fig.~\ref{alig} shows the scheme proposed for aligning the electrodes.

	To reach $10^{-24}\;e\>cm$, the muon current must be increased
500-fold beyond that of the $(g-2)$ measurement.  Therefore the ring
will be converted to strong focussing, and the beam and target systems
upgraded.  A lithium lens is proposed for the latter.  Pions of $900
\> MeV/c$ will be collected so decay muons emitted backward in the
center of mass have the desired $500 \> MeV/c$ momentum and can be
transmitted to the $(g-2)$ ring.  This experiment could use the
maximum intensity listed in Table \ref{t: upgr}.  To reach the
sensitivity goal, about $100 \> TP/1.25$ second cycle is required.
The AGS would be run in single bunch single fast extracted mode at
$13.4 \> GeV/c$.  A total of $10^{15}$ stored muons would be
accumulated.

\section{Conclusion}

	Running the AGS for fixed target physics in the RHIC era is
extremely cost-effective because it can be done on the margin.  The
machine is required only 2-4 hours a day for RHIC and so could
theoretically be available for $> 140$ hours/week for 37 weeks/year.
For modest cost, the AGS slow beam intensity could be tripled, allowing
several interesting new possibilities to become practical.  With only
one or two experiments running at once, properties of the AGS such as
microstructure, primary energy, and duty cycle can be ``customized''
in a way not practical in a large, diverse program.  This would greatly
benefit the proposals discussed above and is a key advantage of AGS-2000.

	The menu of experiments proposed for the AGS-2000 era
include a measurement of $|V_{td}|$, incisive probes of both
Standard Model and non-SM CP-violation, and of a number of low energy
manifestations of supersymmetry, including a type of lepton flavor
violation predicted by GUT-scale SUSY.   These are all 
compelling experiments, and cannot easily be done elsewhere. 

\section{Acknowledgments}

        I would like to thank D. Bryman, I-H. Chiang, P. Cushman,
M. Diwan, J. Frank, S. Kettell, A. Konaka, Y. Kuno, H. Ma,
J.A. Macdonald, W. Molzon, W. Morse, T. Roser, and Y. Semertzidis for
discussions, corrections, plots, access to data and other assistance
with this paper.  This work was supported by the U.S. Department of
Energy under Contract No. DE-AC02-76CH00016.

%\bigskip
%\noindent
%References

%\bibliography

\end{document}